# A unified model for informetrics based on the wave and heat equations

Fred Y. Ye [a, b] Ronald Rousseau [c,d,e]

[a]Dept. Information Resource Management, Zhejiang University, Hangzhou, China
[b]Institute of Scientific and Technical Information of China, Beijing, China
E-mail: yye@zju.edu.cn

[c] KHBO (Association K.U.Leuven), Industrial Sciences and Technology, Zeedijk 101, B-8400 Oostende, Belgium
[d] K.U.Leuven, Dept. Mathematics, Celestijnenlaan 200B, B-3001 Leuven (Heverlee), Belgium
[e] Antwerp University (UA), IOIW, Venusstraat 35, B 2000 Antwerp, Belgium
E-mail: ronald.rousseau@khbo.be

## Abstract

*The function* $g(r,t) = p(r+q)^{-\beta} e^{kt}$ *is introduced as a basic informetric function describing the classical informetric laws (through its Mandelbrot part) and a time evolution. It is shown that this function is a solution of a wave-type and of a heat-type partial differential equation. It is suggested that our approach may lead to a description of informetrics in a partial differential equation setting, formally similar to that for well-known physical laws.*

## Introduction

Mandelbrot's law (Mandelbrot, 1953) states that in a source-item relation the production of the source at rank r is given as:

$$g(r) = \frac{E}{(1+Fr)^{\beta}} \qquad (1)$$

with β > 0, or equivalently

$$g(r) = \frac{p}{(q+r)^{\beta}} \qquad (2)$$

Multiplying the function g(r) by an exponential time-dependent function leads to a function of two variables g(r,t):

$$g(r,t) = \frac{p}{(q+r)^{\beta}} e^{k \cdot t} \qquad (3)$$

When β=0, g(r,t) just describes the time dependency, and when moreover, k=0, g(r,t) reduces to a constant function. Note that, following Egghe (2005) the variable r should not be considered as a natural number, but as a positive real number, so that the function g(r) is a rank-density.

For practical reasons we assume that the Mandelbrot part, as well as the time-dependent part have compact support, i.e. are considered on intervals of the form [1, S] or similar ones. It is well known that the so-called informetric laws such as Bradford's, Lotka's, Mandelbrot's and Leimkuhler's are all mathematically equivalent forms of the same regularity. This equivalence was suggested by Fairthorne (1969), see also (Rousseau, 2005), and shown in different steps. Contributors to prove this equivalence are, among others, Yablonski (1980), Bookstein (1976; 1990), Egghe (1985) and Rousseau (1988). The story and complete proof of these equivalences can be found in Egghe (2005). Models for explaining these laws include the Simon-Price success-breeds-success model (Simon, 1955; Price, 1976; Egghe & Rousseau, 1995), which has recently been re-invented under the name of preferential attachment (Barabási & Albert, 1999) and Mandelbrot's fractal mechanism (Mandelbrot, 1982). Egghe (2005) studies the informetric laws in the setting of Information Production Processes. Recently (Bailón-Moreno et al., 2005) provide another setting based on a set of seven principles. This approach, in which fractals play an essential role, leads to the informetric laws, the law of exponential growth and some ageing principles.

Nowadays power laws, through their relation with the scale-free property (Newman, 2003; Newman, 2005; Egghe, 2005), play an essential role in network studies, such as the Internet, cell networks, aviation networks, business firm networks, pollination networks, disease networks and so on (Boccaletti et al. ,2006; Barrat, 2004; Li et al., 2004; Kitsak et al., 2010; Martin Gonzalez et al., 2010). Besides power law functions also exponential functions play a key role in informetrics, in particular in studying growth over time. In this article we illustrate the potential benefit of using the time-rank function g(r,t), which plays a key role in a simple unified mechanism for describing all these phenomena.

## Fixed time – fixed rank: A theoretical mechanism of informetric laws

When t is fixed (an interesting case is the time origin t = 0), g(r,t) reduces to the Mandelbrot function, which is known to be equivalent to other informetric laws. Moreover, a simple transformation r'= r+1 leads to Zipf's law.

When r is fixed g(r,t) becomes

$$g_t(r) = A \cdot e^{k \cdot t} \qquad (4)$$

which is an increasing (k > 0) or decreasing (k < 0) exponential function. Both have applications in informetric research (Price, 1963; Egghe et al., 1995).

So, using a simple unified informetric function $g(r,t) = p(r+q)^{-\beta} e^{kt}$, we see that g(r, t) becomes Mandelbrot's distribution $g(r) = p(r+q)^{-\beta}$ when k=0 and an exponential time distribution $g(t) = pe^{kt}$ when β = 0, hence combining in a simple way two basic regularities in informetrics.

We recall that, keeping t fixed, the case β=1 already links a power law (in the rank-frequency form) to an exponential (in the size-frequency form), see (Egghe & Rousseau, 2003).

## Differential equations: A unified model for informetrics

Many physical phenomena are described in terms of (partial) differential equations, including information communication (Ye, 1998). We will next show that g(r,t) is a solution of some interesting and well-known partial differential equations. Indeed, the function g(r,t) satisfies the following relationships, involving partial derivatives:

$$\frac{\partial g}{\partial r}(r,t) = -\beta p(q+r)^{-(\beta+1)} e^{k\cdot t} = -\frac{\beta}{q+r} g(r,t) \qquad (5)$$

$$\frac{\partial^2 g}{\partial r^2}(r,t) = \beta(\beta+1) p(q+r)^{-(\beta+2)} e^{k\cdot t} = \frac{\beta(\beta+1)}{(q+r)^2} g(r,t) \qquad (6)$$

$$\frac{\partial g}{\partial t}(r,t) = k g(r,t) \qquad (7)$$

$$\frac{\partial^2 g}{\partial t^2}(r,t) = k^2 g(r,t) \qquad (8)$$

Combining these equations yields:

$$\frac{\partial^2 g}{\partial r^2}(r,t) - \frac{1}{\mathrm{c}^2}\frac{\partial^2 g}{\partial t^2}(r,t) = \left(\frac{\beta(\beta+1)}{(q+r)^2} - \frac{k^2}{c^2}\right) g(r,t) \qquad (9)$$

and

$$\frac{\partial^2 g}{\partial r^2}(r,t) - \frac{1}{\mathrm{d}^2}\frac{\partial g}{\partial t}(r,t) = \left(\frac{\beta(\beta+1)}{(q+r)^2} - \frac{k}{d^2}\right) g(r,t) \qquad (10)$$

where c and d are arbitrary non-zero constants.

Setting

$$\mathrm{a}(r) = \left(\frac{\beta(\beta+1)}{(q+r)^2} - \frac{k^2}{c^2}\right) \qquad (11)$$

and

$$b(r) = \left(\frac{\beta(\beta+1)}{(q+r)^2} - \frac{k}{d^2}\right) \qquad (12)$$

yields:

$$\frac{\partial^2 g}{\partial r^2}(r,t) - \frac{1}{\mathrm{c}^2}\frac{\partial^2 g}{\partial t^2}(r,t) = a(r) g(r,t) \qquad (13)$$

and

$$\frac{\partial^2 g}{\partial r^2}(r,t) - \frac{1}{\mathrm{c}^2}\frac{\partial g}{\partial t}(r,t) = b(r) g(r,t) \qquad (14)$$

This shows that the function g(r,t) is a solution of these two partial differential equations. These two partial differential equations are essentially a wave-type equation and a heat-type equation, but with a non-constant coefficient.

## Fundamental solutions and the function g(r,t) as a fundamental solution

Let *L*(*D*) be a differential operator with constant coefficients. Then a fundamental solution (Renardy and Rogers, 2004) for *L* is a function G (more generally a distribution or generalized function) satisfying the equation $L(D)G = \delta$, where δ is the Dirac δ function. Of course, fundamental solutions are unique only up to a solution of the homogeneous equation $L(D)u = 0$. The significance of the fundamental solution lies in the fact that $L(D)(G * f) = (L(D)G) * f = \delta * f = f$, provided that the convolution $G * f$ is defined. Recall (Rousseau, 1998) (where the importance of convolutions in information science is illustrated) that a convolution of two real functions g and h is defined as

$$(g * h)(t) = \int_{-\infty}^{+\infty} g(t-u)h(u)\,du \qquad (15)$$

For functions of two variables this becomes:

$$(g * s)(r,t) = \iint_{R^2} g(r-u,t-v)s(u,v,)dudv \qquad (16)$$

We have shown that g(r,t) is a solution of a wave-type and of a heat-type partial differential equation. In order to be a fundamental solution we must have constants coefficients. This leads to the additional requirements:

$$\frac{\beta(\beta+1)}{(r+q)^2} = \frac{k^2}{c^2} = \frac{k}{d^2} \qquad (17)$$

## The time-type distribution of the unified informetric model

Taking r constant, and subtracting (13) from (14) yields (after multiplication with $c^2$):

$$\frac{d^2g}{dt^2}-\frac{c^2}{d^2}\frac{dg}{dt}=c^2(b-a)g(t) \qquad (18)$$

where a and b are now constants.

The characteristic equation of this homogeneous linear differential equation with constant coefficients is $r^2-(c^2/d^2)r-c^2(b-a)=0$. Its two solutions are $r_{1,2}=[(c^2/d^2)\pm\sqrt{(c^4/d^4)+4c^2(b-a)}]/2$ (assuming that we have two different solutions) so that the general solution of Eq. (18) becomes

$$g(t)=c_1e^{r_1 t}+c_2e^{-r_2 t} \qquad (19)$$

(which can be rewritten using sines and cosines in case the roots are complex conjugate). In the special case that $c^2 = 4d^4(a-b)$ the characteristic equation has a double root equal to $c^2/d^2$ and

$$g(t)=C_3e^{(c^2/d^2)t}+C_4t\,e^{(c^2/d^2)t} \qquad (20)$$

**Discussion and conclusions**

We propose a generalized informetric theory based on the wave and heat equations and raise the question about which functions a(r) and b(r) lead to meaningful solutions and which are the domains on which these functions must be studied. Clearly the function $g(r,t)=\frac{p}{(q+r)^{\beta}}e^{k\cdot t}$ is an interesting common solution of the partial differential equations (13) and (14). Are there other solutions? How can they be described?

General solutions f(r, t) fitting the partial differential equations (21) and (22):

$$\frac{\partial^2 f}{\partial r^2}-\frac{1}{c^2}\frac{\partial^2 f}{\partial t^2}=a(r)g(r,t)=s_1(r,t) \qquad (21)$$

$$\frac{\partial^2 f}{\partial r^2}-\frac{1}{d^2}\frac{\partial f}{\partial t}=b(r)g(r,t)=s_2(r,t) \qquad (22)$$

introduce two branches:

$$f_1(r,t) = f_1[g(r,t)] = (g * s_1)(r,t) = \iint_{R^n} g(r-u, t-v) s_1(u,v,) du dv \quad (23)$$

$$f_2(r,t) = f_2[g(r,t)] = (g * s_2)(r,t) = \iint_{R^n} g(r-u, t-v) s_2(u,v,) du dv \quad (24)$$

in which g(r, t) is a core function, playing a similar role as the heat-core in the heat equation. We expect the combined wave and heat equations to become a unified research framework. Its study may produce rich new phenomena.

We conclude that in view of the results of this contribution the wave-heat equations lead to a framework to study [time-dependent] informetric laws in terms of partial differential equations, formally similar to that for well-known physical laws. We would, moreover, like to point out that the wave and the heat equation describe physical phenomena, while our informetric framework uses a generalized wave-type and heat-type set of equations.

## Acknowledgements

The authors thank the financial support of the National Natural Science Foundation of China (NSFC Grant No 70773101) for supporting their collaboration. R.R. thanks his colleague Ye Ying for his hospitality at the Department of Information Resource Management, Zhejiang University, where most of this work has been done.